\documentstyle[preprint,aps,eqsecnum,axodraw]{revtex}

\newcommand{\beq}{\begin{equation}}
\newcommand{\eeq}{\end{equation}}
\newcommand{\snu}{\tilde \nu}

\newcommand{\msnuone}{m_{\snu_+}}
\newcommand{\msnutwo}{m_{\snu_-}}

\newcommand{\dmsnums}{{\mbox{$\Delta m_{\tilde \nu_m}^2$}}}
\newcommand{\dmsnutwo}{{\mbox{$\Delta m^2_{\snu}$}}}

\newcommand{\Sym}{\mbox{\rm Sym}_2}

\newcommand{\Tr}{{\mbox{Tr}}}
\newcommand{\Cof}{{\mbox{cof}}}
\newenvironment{Eqnarray}%
         {\arraycolsep 0.14em\begin{eqnarray}}{\end{eqnarray}}
\def\beqa{\begin{Eqnarray}}
\def\eeqa{\end{Eqnarray}}
\def\mz{m_Z}
\def\hh{H^0}
\def\hl{h^0}
\def\ha{A^0}
\def\mhh{m_{\hh}}
\def\mhl{m_{\hl}}
\def\mha{m_{\ha}}
\def\msnusnuab{(M^2_{\snu\snu^*})_{\alpha\beta}}
\def\msnusnuij{(M^2_{\snu\snu^*})_{ij}}
\def\msnusnu{M^2_{\snu\snu^*}}
\def\msnuiv{M^4_{\snu\snu^*}}

\def\l{\lambda}

\def\lp{\lambda^\prime}

\def\ifmath#1{\relax\ifmmode #1\else $#1$\fi}
\def\eighth{\ifmath{{\textstyle{1 \over 8}}}}

\def\quarter{\ifmath{{\textstyle{1 \over 4}}}}

\def\refs#1#2{refs.~\cite{#1} and \cite{#2}}
\def\Ref#1{ref.~\cite{#1}}

\def\eq#1{eq.~(\ref{#1})}
\def\Eq#1{Eq.~(\ref{#1})}
\def\eqs#1#2{eqs.~(\ref{#1}) and (\ref{#2})}

\def\vev#1{\langle{#1}\rangle}
\def\cross{\times}
\def\npb#1{{\sl Nucl.\ Phys.}\ {\bf B#1}}
\def\plb#1{{\sl Phys.\ Lett.}\ {\bf B#1}}
\def\prd#1{{\sl Phys.\ Rev.}\ {\bf D#1}}
\def\prl#1{{\sl Phys.\ Rev.\ Lett.} {\bf #1}}

\def\epjc#1{{\sl Eur.~Phys.~J.}\ {\bf C#1}}

\def\ifmath#1{\relax\ifmmode #1\else $#1$\fi}
\def\half{\ifmath{{\textstyle{1 \over 2}}}}

\def\eighth{\ifmath{{\textstyle{1 \over 8}}}}

\begin{document}

\draft{\tighten

\preprint{
\vbox{
      \hbox{SLAC-PUB-8464}
      \hbox{SCIPP-00/17}
      \hbox{hep-ph/0005276}
      \hbox{May 2000}
    }}

\bigskip
\bigskip

\renewcommand{\thefootnote}{\fnsymbol{footnote}}

\title{Basis-independent analysis of the sneutrino sector in R-parity 
violating supersymmetry} 
\footnotetext{Research supported
by the Department of Energy under contracts DE-AC03-76SF00515
and DE-FG03-92ER40689.} 
\author{Yuval Grossman\,$^a$ and  Howard E. Haber\,$^b$}
\address{
$^a$Stanford Linear Accelerator Center, 
        Stanford University, Stanford, CA 94309 \\
  $^b$Santa Cruz Institute for Particle Physics, 
University of California, Santa Cruz, CA 95064}

\maketitle

\begin{abstract}%
In R-parity-violating supersymmetric models (with a conserved baryon
number), there are no quantum numbers that distinguish
the lepton-doublet and down-type Higgs supermultiplets. 
As a result, the R-parity-violating parameters depend
on the basis choice for these superfields, although physical observables
are independent of the choice of basis.  This paper presents a 
basis-independent computation of the sneutrino/antisneutrino
squared-mass splitting in terms of basis-independent
quantities.  Techniques are developed for an arbitrary number of sneutrino
generations; specific results are provided for the one, two and three
generation cases.

\end{abstract}

}

\newpage

\renewcommand{\thefootnote}{\alph{footnote}}

\section{Introduction}
In low-energy supersymmetric extensions of the Standard Model, lepton
and baryon number conservation are not automatically respected by the
most general set of renormalizable interactions \cite{dreinerreview}.  
Nevertheless,
experimental observations imply that lepton number violating effects,
if they exist, must be rather small.  Moreover, baryon number
violation, if present, must be consistent with the observed stability
of the proton.  If one wants to enforce lepton and baryon
number conservation, it is sufficient to impose one extra discrete
symmetry.  In the minimal supersymmetric extension of the Standard
Model (MSSM), a multiplicative symmetry called R-parity is 
introduced \cite{fayet},
such that the R quantum number of an MSSM field of spin $S$, baryon
number $B$ and lepton number $L$ is given by $(-1)^{[3(B-L)+2S]}$. By
introducing \hbox{$B\!\!-\!\!L$} conservation modulo 2, one eliminates
all dimension-four lepton number and baryon number-violating
interactions.

The observation of neutrino mixing effects in solar and
atmospheric \cite{pdg} 
neutrinos suggest that lepton-number is not an exact global
symmetry of the low-energy theory.  One can develop a supersymmetric
model of neutrino masses that generalizes the see-saw mechanism while
maintaining R-parity as a good symmetry \cite{susyseesaw,GH} 
(where lepton number is violated by two units).
In this paper, we consider the alternative possibility that neutrino
masses and mixing arise in a theory of R-parity violation, in which
lepton number is violated by one unit \cite{all}.  In the most general
R-parity-violating (RPV) model, both $B$ and $L$ are violated.
However, it is difficult to relax
both lepton and baryon number conservation in the low-energy theory
without generating a proton decay rate many orders of magnitude above
the present bounds.  It is possible to enforce baryon number
conservation, while allowing for lepton number violating interactions
by imposing a discrete baryon ${\bf Z_3}$ symmetry on the low-energy
theory \cite{IbRo}, 
in place of the standard ${\bf Z_2}$ R-parity.  Henceforth,
we consider R-parity-violating low-energy supersymmetry
with an unbroken discrete baryon ${\bf Z_3}$ symmetry.  
This model exhibits
lepton-number-violating phenomena such as neutrino masses,
sneutrino/antisneutrino mixing, and lepton-number violating decays.

In RPV low-energy supersymmetry, there is no 
quantum number that distinguishes the lepton supermultiplets $\hat
L_m$ and the down-type Higgs supermultiplet $\hat H_D$
($m$ is a generation label that runs from 1 to $n_g=3$).  Each
supermultiplet transforms as a $Y=-1$ weak doublet under the electroweak
gauge group.  It is therefore convenient to denote $\hat L_0\equiv \hat H_D$
and unify the four supermultiplets by one symbol $\hat L_\alpha$
($\alpha=0,1,\ldots,n_g$). 
Then, the relevant terms in the (renormalizable) superpotential are
\beq \label{rpvsuppot}
W=\epsilon_{ij} \left[
-\mu_\alpha \hat L_\alpha^i \hat H_U^j + 
\half\l_{\alpha\beta m}\hat L_\alpha^i \hat L_\beta^j \hat E_m +
\lp_{\alpha nm} \hat L_\alpha^i \hat Q_n^j  \hat D_m
\right]\,,
\eeq
where $\hat H_U$ is the up-type Higgs supermultiplet, the
$\hat Q_n$ are doublet quark supermultiplets, the $\hat D_m$ are 
singlet down-type quark supermultiplets
and the $\hat E_m$ are the singlet charged lepton 
supermultiplets. 
Note that $\mu_\alpha$ and $\lp_{\alpha nm}$ are vectors 
and $\l_{\alpha\beta m}$ is an antisymmetric matrix in 
the generalized lepton flavor space.

Next, the
soft-supersymmetry-breaking terms are also generalized in similar way.
The relevant terms are
\beq \label{softsusy}
 V_{\rm soft}  =
  (M^2_{\widetilde L})_{\alpha\beta}\,
          \widetilde L^{i*}_\alpha\widetilde L^i_\beta
  -(\epsilon_{ij} b_\alpha\tilde L_\alpha^i H_U^j +{\rm h.c.}) 
 +  \epsilon_{ij} \bigl[\half a_{\alpha\beta m} \widetilde L^i_\alpha
       \widetilde L^j_\beta \widetilde E_m + a'_{\alpha nm}
       \widetilde L^i_\alpha\widetilde Q^j_n\widetilde D_m + {\rm h.c.}\bigr]
 \,,
\eeq
where the fields appearing in \eq{softsusy} are the scalar partners of
the superfields that appear in \eq{rpvsuppot}.
Here, $b_\alpha$ and $a'_{\alpha nm}$ are vectors, 
$a_{\alpha\beta m}$ is an antisymmetric matrix and 
$(M^2_{\widetilde L})_{\alpha\beta}$ is a Hermitian matrix
in the generalized lepton flavor space.

When the scalar potential is minimized (see Section~II), one finds
a vacuum expectation value for the neutral scalar fields:
$\vev{\tilde L_\alpha}=v_\alpha/\sqrt{2}$ 
and $\vev{H_U}=v_u/\sqrt{2}$.
To make contact with the usual notation of the MSSM, we define the
length of the vector $v_\alpha$ by $v_d \equiv (v_\alpha v_\alpha)^{1/2}$ and
$\tan\beta \equiv v_u/v_d$.  The mass of the $W$ boson constrains
the value $v^2 \equiv v_u^2+v_d^2=(246~{\rm GeV})^2$.

So far, there is no distinction between the neutral Higgs bosons and
neutral sleptons.  Nevertheless, we know that RPV-interactions, if
present, must be small.  It is tempting to choose a
particular convention corresponding to a specific choice of basis
in the generalized lepton flavor space.  For example, one can choose
to {\it define} the down-type Higgs multiplet such that
$\vev{H_D}\equiv\vev{\tilde L_0}=v_d/\sqrt{2}$ and
$\vev{\tilde L_m}=0$.  This means that we let the dynamics 
(which determines the direction of the 
vacuum expectation value in the generalized lepton
flavor space) choose the definition of the down-type Higgs field. 
In this basis, all the RPV-parameters are well defined and must be
small to satisfy phenomenological constraints.

Nevertheless, the above convention is only one possible basis choice.
Other conventions are equally sensible.  For example, one could choose
a second basis where $\mu_m=0$ and a third basis where $b_m=0$.  In
each case, the corresponding RPV-parameters are small.  But comparing
results obtained in different bases requires some care.  Moreover,
it is often desirable to study the evolution of couplings from some
high (unification) scale to the low-energy (electroweak) scale.  The
renormalization group equations for the RPV-parameters is not basis
preserving.  That is, a particular basis choice at the high energy
scale will lead to some complicated effective basis choice at the
low-energy scale.

The problems described above can be ameliorated by avoiding
basis-specific definitions of parameters.  The challenge of such an
approach is to determine a set of basis-independent RPV
parameters, in the spirit of the Jarlskog invariant which characterizes
the strength of CP-violation in the Standard Model \cite{Jar}.  
Such an approach
has been applied to RPV models in the past, where neutrino masses
\cite{bgnn,enrico,Fer,pol,losada}, early universe physics \cite{Davi} and the
Higgs sector \cite{DLR} were studied.  It is instructive to examine
the neutrino spectrum of the RPV model.
At tree level, one neutrino
become massive due to the RPV mixing of the neutrinos and the
neutralinos.  The other $n_g-1$ neutrinos remain massless at
tree-level, although they can acquire smaller radiative masses at
one-loop. 
To first order in the small RPV-parameter the basis
independent formula can be written in the following form \cite{enrico,GHrpv}: 
\beq
\label{numass} 
m_\nu = {m_Z^2 \mu M_{\tilde \gamma}\cos^2\beta \over
m_Z^2 M_{\tilde \gamma}\sin 2\beta-M_1 M_2 \mu} |\hat v \times
\hat\mu|^2\,, 
\eeq 
where $M_{\tilde \gamma}\equiv \cos^2\theta_W M_1 +
\sin^2\theta_W M_2$ depends on gaugino mass parameters $M_1$ and $M_2$.
In \eq{numass}, $\hat v$ and $\hat\mu$ are unit vectors in the
$v_\alpha$ and $\mu_\alpha$ directions, respectively.  It is
convenient to introduce the notation of the cross-product of two
vectors.  Although the cross-product technically exists only in
three-dimensions, the dot product of two cross-products can be
expressed as a product of dot-products 
\beq \label{xproduct}
(a\cross b)\cdot(c\cross d) = (a \cdot c)(b \cdot d)-(a \cdot d)(b \cdot c)\,, 
\eeq 
which exists in any number of dimensions.  This notation is useful, since
any expression that involves
the cross-product of two vectors vanishes if the 
corresponding vectors are parallel.
This provides a nice geometrical characterization of the small
RPV-parameters of the model.  For example, $|\hat v \cross
\hat\mu|^2=\sin^2\xi$ where $\xi$ is the angle between $\hat v$ and
$\hat\mu$.  Thus, in \eq{numass},  $|\hat v \times \hat\mu|^2$ is the small
RPV-parameter, while the prefactor can be computed in the
R-parity-conserving (RPC) limit of the model.

In this paper, we focus on a basis-independent description of the
RPV-parameters that govern the sneutrino spectrum.
The model possesses lepton-number-violating
$\Delta L=1$ interactions that give rise to the mixing of
sleptons and Higgs bosons.\footnote{These interactions modify the
phenomenology of the charged and neutral scalars (relative to the
RPC limit), as discussed in \Ref{DLR}.}     
These interactions also generate $\Delta L=2$ effective operators that
give rise to sneutrino/antisneutrino mixing
\cite{GH,losada,GHrpv,HKK,HMM,FGH,Bar-Shalom,davking,dpftalk}.  
In this case, the sneutrino ($\snu$)
and antisneutrino ($\bar{\snu}$), which are eigenstates of lepton
number, are no longer mass eigenstates.  The mass eigenstates are
superpositions of $\snu$ and $\bar{\snu}$, and sneutrino mixing
effects can lead to a phenomenology analogous to that of
$K$--$\overline K$ and $B$--$\overline B$ mixing \cite{GH}.  The mass splitting
between the two sneutrino mass eigenstates is related to the magnitude
of lepton number violation, which is typically characterized by the
size of neutrino masses \cite{GH,HMM}.  As a result, the
sneutrino/antisneutrino mass splitting is expected generally to be
very small. Yet, it can be detected in many cases, if one is able to
observe the lepton number oscillation \cite{GH}.

In contrast to the neutrino sector (where only one neutrino mass eigenstate
acquires a tree-level mass), in general {\it all}
sneutrinos/antisneutrino pairs are split in mass at tree 
level.\footnote{This result is a consequence of the fact that
in the RPC limit, neutrinos are massless and hence degenerate, whereas
the RPC sneutrino masses are in general non-degenerate.  See Section V
and \Ref{GHrpv} for further discussions of this point.}
For simplicity, we consider the case of a CP-conserving scalar
sector.  In the RPC limit,
the CP-even scalar sector consists of two
Higgs scalars ($\hl$ and $\hh$, with $\mhl<\mhh$) and $n_g$
generations of CP-even sneutrinos
$(\snu_+)_m$, while the CP-odd scalar sector consists of 
the Higgs scalar, $\ha$, the Goldstone
boson (which is absorbed by the $Z$), and $n_g$ generations of
CP-odd sneutrinos $(\snu_-)_m$.  Here, we have implicitly chosen a
flavor basis in which the sneutrinos are mass eigenstates.
Moreover, the $(\snu_\pm)_m$ are mass
degenerate (separately for each $m$),
so that the standard practice is to define eigenstates of
lepton number: $\snu_m\equiv [(\snu_+)_m + i(\snu_-)_m]/\sqrt{2}$ and 
$\overline{\snu}_m\equiv\snu_m^*$.  When R-parity is violated, the sneutrinos
in each CP-sector mix with the corresponding Higgs scalars, and the
mass degeneracy of $(\snu_+)_m$ and $(\snu_-)_m$ is broken.
In \Ref{GHrpv} we computed the mass-splitting in a special basis where 
$v_m=0$ 
and the matrix $\msnusnuij$ 
[which is the $3\times 3$ block sub-matrix of
$\msnusnuab$ defined in \eq{treelevelsnumasses}] is diagonal.
In this basis, we identified the $b_m$ as the relevant
small RPV-parameters.   To leading
order in $b_m^2$, 
\beq \label{dms}
(\dmsnutwo)_m = {-4 \, b_m^2\, m_Z^2 \, m_{\snu_m}^2 \, \sin^2\beta \over
(m_H^2-m_{\snu_m}^2) (m_h^2-m_{\snu_m}^2) (m_A^2-m_{\snu_m}^2)}\,,
\eeq
where $(\dmsnutwo)_m \equiv  (\msnuone^2)_m- (\msnutwo^2)_m$.  As in
the neutrino case described above, we may evaluate the prefactor that
multiplies $b_m^2$ in the RPC limit.  In deriving
\eq{dms}, it was assumed that all RPC Higgs and
sneutrino masses are all distinct.  If degeneracies exist, the above
formula must be modified.

The goal of this paper is to reanalyze the sneutrino mass spectrum in
a basis-independent formalism.  We identify the small RPV-parameters
that govern the sneutrino/antisneutrino mass splittings.  Our
technique will also allow us to generalize the analysis to treat the
case of scalar mass degeneracies.  In Section~II, we derive a
convenient form for the CP-even and CP-odd scalar squared-mass
matrices.  We compute the sneutrino/antisneutrino squared-mass
difference in the case of one sneutrino flavor in Section~III.
In Section~IV, we generalize to an arbitrary number of generations,
and exhibit explicit formulae for the two and three generation cases.  
The latter results assume that in the RPC limit, there are no
degeneracies among different sneutrino flavors.  The degenerate case
is treated in Section~V.  A discussion of our results and conclusions
are presented in Section~VI.  Details of our computations are provided
in six appendices.

\section{Minimum condition and basic equations}

We begin our analysis by collecting the relevant formulae given 
in \Ref{GHrpv}.  We assume that the scalar sector is 
CP-conserving,\footnote{In the MSSM, the Higgs sector is
automatically CP-conserving at tree-level, since all phases
can be removed by suitable redefinitions of the fields.  In the RPV
model, new phases enter through $(M^2_{\tilde L})_{\alpha\beta}$,
$b_\alpha$ and $\mu_\alpha$, which cannot all be 
simultaneously removed in the general case.}
which implies that the scalar fields can be defined such that
$M^2_{\tilde L}$ is a real symmetric matrix and $b_\alpha$
and $\mu_\alpha$ are real.
The vacuum expectation value $\vev{L_\alpha}\equiv v_\alpha/\sqrt{2}$ is
determined by minimizing the scalar potential.  With the assumption of
CP-conservation, one can separate out the scalar
potential for the CP-even and CP-odd sector, $V=V_{\rm even}+V_{\rm odd}$.
Then, $v_\alpha$ is determined 
by minimizing $V_{\rm even}$, and the resulting condition is given by:
\beq \label{veveq}
\msnusnuab v_\beta=v_u b_\alpha\,,
\eeq
where
\beq \label{treelevelsnumasses}
\msnusnuab\equiv
({M^2_{\tilde L}})_{\alpha\beta}+ \mu_\alpha\mu_\beta-\eighth 
(g^2+g^{\prime 2})(v_u^2-v_d^2)\delta_{\alpha\beta}\,.
\eeq
Note that \eq{veveq} determines both the size of $v_\alpha$ and
its direction.  In a perturbative treatment of RPV terms,
we can use the RPC value for the squared magnitude 
$v_d^2\equiv\sum_\alpha
v_\alpha v_\alpha$, and then use \eq{veveq} to determine the
direction of $v_\alpha$ in the generalized lepton flavor space.

We next separate the scalar squared-mass matrices into CP-odd and
CP-even blocks.  In the $H_U$--$\widetilde L_\alpha$ basis, the
CP-odd squared-mass matrix is given by
\beq \label{modd2}
M_{\rm odd}^2= \pmatrix{
b_\rho v_\rho/v_u & b_\beta \cr
b_\alpha & \msnusnuab \cr}\,,
\eeq
while the CP-even squared-mass matrix is given by
\beq \label{meven2}
M_{\rm even}^2= \pmatrix{
\quarter(g^2+g^{\prime 2})v_u^2+b_\rho v_\rho/v_u&
-\quarter(g^2+g^{\prime 2})v_u v_\beta -b_\beta\cr
-\quarter(g^2+g^{\prime 2})v_u v_\alpha -b_\alpha &
\quarter(g^2+g^{\prime2})v_\alpha v_\beta +\msnusnuab\cr}\,,
\eeq
where $\msnusnuab$ is defined in \eq{treelevelsnumasses}.

To compute the squared-mass differences of the corresponding 
CP-even and CP-odd sneutrinos, we must diagonalize both of the
above matrices.  We wish to employ a perturbative procedure by
identifying the small RPV-parameters, without resorting to a specific
choice of basis.  Our strategy is to recast the two scalar squared-mass
matrices in a more convenient form.  

First, consider the CP-odd squared mass matrix [\eq{modd2}].
Note that the vector $(-v_u,v_\beta)$ is an eigenvector of $M_{\rm
odd}^2$ with zero eigenvalue; this is the Goldstone boson that is
absorbed by the $Z$.  We can remove the Goldstone boson by introducing
the following orthogonal $(n_g+2)\times (n_g+2)$ matrix
\beq
U_{o}=\pmatrix{ -v_u/v & v_d/v & 0 \cr
v_\beta/v & v_u v_\beta/(v_d v)  & X_{\beta i}}\,, 
\eeq
where $v\equiv (v_u^2+v_d^2)^{1/2}$.
Note that the index $i$ runs from 1 to $n_g$; thus $X_{\alpha i}$ is
an $(n_g+1) \times n_g$ matrix. The orthogonality of $U_o$ implies
that each column of $U_o$ is a real unit vector and different columns
are orthogonal. In addition, the set $\{v_\beta/v,X_{\beta i}\}$
forms an orthonormal set of vectors in an $(n_g+1)$-dimensional vector
space.  It follows that:
\beqa 
v_\alpha X_{\alpha i} &=& 0\,, \label{ort1} \\
X_{\alpha i}X_{\alpha j} &=& \delta_{ij}\,, \label{ort2} \\
X_{\alpha i}X_{\beta i} &=& \delta_{\alpha \beta} 
   - {v_\alpha v_\beta \over v_d^2}\,. \label{ort3} 
\eeqa
In our computations, no explicit realization of the $X_{i\alpha}$ will
be required.  A simple computation yields:
\beq \label{modd2d1}
U_o^T M_{\rm odd}^2 U_o = \pmatrix{0 & 0_\beta \cr 0_\alpha & 
(\widetilde M_{\rm odd}^2)_{\alpha\beta} }\,,
\eeq
where $0_\beta$ [$0_\alpha$] is a row [column] matrix of zeros 
and\footnote{We define $X_{i\alpha}$ to be the
transpose of the matrix $X_{\alpha i}$.  When no ambiguity arises, we
will not explicitly exhibit the transpose symbol (superscript $T$).}
\beq \label{modd2d2}
\widetilde M_{\rm odd}^2= \pmatrix{
v^2 (v \cdot b)/(v_u v_d^2) & 
v b_\beta X_{\beta i}/v_d \cr
v X_{j\alpha}b_\alpha/v_d & 
X_{j \alpha} \msnusnuab X_{\beta i}
}\,, 
\eeq
where $v\cdot b\equiv v_\alpha b_\alpha$.
The eigenstates of $\widetilde M_{\rm odd}$ correspond to the
CP-odd Higgs boson $\ha$ and $n_g$ generations of CP-odd sneutrinos.

It turns out that it is also convenient to rotate the CP-even
squared-mass matrix, but by a slightly different orthogonal
transformation.  In the limit of $\mz=0$, the CP-even squared-mass
matrix also possesses a Goldstone boson, which we can explicitly
isolate.  Comparing the CP-odd and CP-even cases, we see that when
$g=g'=0$ the two matrices are related by $b_\alpha\to -b_\alpha$ and
$v_u\to -v_u$.  Thus, if we introduce $U_e\equiv U_o(v_u\to -v_u)$
and define $\widetilde M^2_{\rm even}\equiv U_e^T M_{\rm even}^2 U_e$, then
\beq \label{meven2d}
\widetilde M_{\rm even}^2= \pmatrix{m_Z^2 \cos^2 2\beta & 
-m_Z^2 \cos 2\beta \sin2\beta& 0 \cr
-m_Z^2 \cos 2\beta \sin2\beta & 
m_Z^2 \sin^2 2\beta +  v^2 (v\cdot b)/(v_u v_d^2) & 
-v b_\beta X_{\beta i}/v_d \cr
0 & -v X_{j \alpha}b_\alpha/v_d & 
X_{j \alpha} \msnusnuab X_{\beta i}
}\,, 
\eeq
where we used $m_Z^2=\quarter(g^2+g^{\prime 2})v^2$ and
$\tan\beta\equiv v_u/v_d$.  
The eigenstates of $\widetilde M_{\rm even}^2$ correspond to the
CP-even Higgs bosons $\hl$ and $\hh$ 
and $n_g$ generations of CP-even sneutrinos.

In the RPC limit, one can choose a basis in which 
$v_m=b_m=X_{0m}=0$.  It follows
that the sneutrino and Higgs mass matrices decouple and one
recovers the known RPC result.  
A basis-independent characterization of the RPC
limit in the scalar (sneutrino/Higgs) sector
is the condition that the vectors $b_\beta$ and $v_\beta$ are
aligned.\footnote{For example, if one chooses the basis where
$v_m=0$, then by \eq{ort1}, $X_{0m}=0$.  In this basis, the $b_m$ are the
small RPV-parameters.}  
Equivalently, by using the transformed mass
matrices given above, it is clear that the quantities
\beq \label{capbee}
B_i\equiv {vb_\beta X_{\beta i}\over v_d}  
\eeq
can be identified as $n_g$ basis-independent parameters that vanish in
the RPC limit, and thus provide good candidates
for the small quantities that can be used in a perturbative expansion.
If $b_\beta$ and $v_\beta$ are aligned, then \eq{ort1} implies that
the $B_i=0$ and we are back to the RPC limit.

Although the $B_i$ provide a basis-independent set of small RPV-parameters,
the explicit dependence on $X_{\alpha i}$ is inconvenient.
Clearly, it is preferable to re-express the $B_i$ directly in terms of the
original model parameters.  In the following sections, we will exhibit
this procedure in the one-generation case and generalize it to the
multi-generation case.   To this end, it is convenient to introduce a
set of new RPV-parameters.  In the case of $n_g\leq 3$, only one new
vector is required for our final results:
\beq \label{cdef}
c_\alpha \equiv {\msnusnuab b_\beta\over b^2}\,.
\eeq
For $n_g>3$, further vectors are required.  It is convenient to introduce a
series of vectors 
\beq \label{cseries}
c^{(n+1)}_\alpha\equiv \msnusnuab c^{(n)}_\beta\,,
\eeq
where $c^{(1)}_\alpha\equiv c_\alpha$.  Clearly, the maximal number of
linearly independent vectors (along with $b$ and $v$) is $n=n_g-1$,
although not all of these will appear in our final results.

To simplify the presentation we introduce the following shorthand for
the elements of the transformed CP-even and CP-odd squared-mass matrices:
\beq \label{modd2dnot}
\widetilde M_{\rm odd}^2= \pmatrix{A & B_i \cr B_j & C_{ij}}\,,
\eeq
\beq \label{meven2dnot}
\widetilde M_{\rm even}^2= 
\pmatrix{D & E &0 \cr E& F+A & -B_i \cr 0 & -B_j  & C_{ij}}\,.
\eeq
Our strategy is to employ a first-order perturbative analysis to
diagonalize these matrices, taking the small parameters to be the
$B_i$ [\eq{capbee}].  At zeroth order (setting the $B_i$ to zero), the
above matrices can be evaluated in the RPC limit.   In particular,
$A=m_A$, the eigenvalues of $\left({D\atop E}\,{E\atop F+A}\right)$ are the
CP-even Higgs squared-masses $\mhl^2$ and $\mhh^2$, and $C_{ij}$ is
the RPC sneutrino squared-mass matrix.  In order for non-degenerate
perturbation theory to be valid, we henceforth assume that none of the
eigenvalues of $C_{ij}$ is (approximately) equal to any of the 
neutral Higgs boson 
squared-masses.\footnote{Relaxing this assumption would require
one to use degenerate perturbation theory.  The resulting analysis
would be more involved and not very illuminating, so we spare
the reader by omitting the consideration of this possibility.}
This is not a serious restriction, since there is no
natural choice of supersymmetric parameters that would guarantee
such a (near) degeneracy.  In particular, the Higgs mass matrices 
arise in part from the mixing of $H_D$ and $H_U$, which is 
governed by the parameter $b_0$ and not 
related to the generation of sneutrino masses.  

The result of the calculation outlined above will be basis-independent
expressions for the sneutrino/antisneutrino squared-mass splittings.
We first illustrate the method for one sneutrino generation in
Section~III, and then generalize it to the multi-generational case in
Section~IV, assuming that the eigenvalues of $C_{ij}$ are
non-degenerate.  The case where some of the eigenvalues of
$C_{ij}$ are degenerate will be treated in Section V.

\section{The case of one generation}
In the one generation case we can drop the Roman indices 
from  $X$, $B$ and $C$, namely we define $X_\alpha\equiv X_{\alpha 1}$, 
$B\equiv B_1$ and $C \equiv C_{11}$. 
To zeroth order in $B$, the CP-even and CP-odd sneutrino squared-masses are
equal to $C$. The corrections can be calculated perturbatively.

The eigenvalue equation for the CP-odd squared-mass matrix,
$\det(\widetilde M_{\rm odd}^2-\lambda I)=0$ reads
\beq
(A-\lambda)(C-\lambda) - B^2=0\,.
\eeq
For $B$ small, $\lambda=C+{\cal O}(B^2)$, so we can take 
$\lambda-C = a B^2$ and solve for $a$.  Thus, to first order in $B^2$,
the squared mass of the CP-odd sneutrino is
\beq
m^2_{\rm odd}= C-{B^2 \over A-C}\,.
\eeq
A similar analysis for the CP-even squared-mass matrix yields
the following result for the squared-mass of the CP-even sneutrino:
\beq
m^2_{\rm even}= C-{B^2 (D-C)\over (F+A-C)(D-C)-E^2}\,.
\eeq
The squared-mass splitting, $\dmsnutwo = m^2_{\rm even}-m^2_{\rm odd}$,
is given by  
\beq \label{sqmassdiff}
\dmsnutwo = {-FCB^2 \over (A-C)[(F+A-C)(D-C)-E^2]}\,,
\eeq
where we have used the fact that $FD=E^2$.  Note that 
the only small parameter in \eq{sqmassdiff} is $B^2$.  We can therefore 
use the RPC values for the prefactor that multiplies $B^2$.
For example, as noted above, $C=m_{\snu}^2$ is the RPC sneutrino squared-mass.

Although $B=vb_\alpha X_\alpha/v_d^2$ is expressed in a basis-independent
manner, the explicit dependence on $X_\alpha$ is inconvenient.
Clearly, it is preferable to re-express $B$ directly in terms of the
original model parameters.
To this end, note that the orthogonality condition [\eq{ort1}]
implies that $X_\alpha$
and $v_\alpha$ are orthogonal, and thus a dot product of any vector 
with $X$ is equivalent to a cross product with $v$. Using \eq{app1} 
for $B^2$ and the RPC values for the
other parameters of \eq{sqmassdiff}, we end up with
\beq \label{sqmassdiff1}
\dmsnutwo = 
{-4 \, b^2 m_Z^2 \, m_{\snu}^2 \, \sin^2\beta \over
(\mhh^2-m_{\snu}^2) (\mhl^2-m_{\snu}^2) (\mha^2-m_{\snu}^2)} 
|\hat v \cross \hat b|^2 
\,,
\eeq
where $\hat b$ is a unit vector in the $b_\alpha$ direction and the
square of the cross-product is formally defined according to
\eq{xproduct}.  
It is easy to check [see Appendix~B] that in the special basis where $v_1=0$,
the basis-independent result above [\eq{sqmassdiff1}] reduces to the 
basis-dependent result quoted in \eq{dms}.

The basis-independent result obtain in \eq{sqmassdiff1} is still not
in optimal form, since it depends on $v$, which is a derived quantity
that requires one to determine the minimum of the scalar potential
[\eq{veveq}].
However, we can employ the vector $c_\alpha$ [defined in \eq{cdef}] to
our advantage by noting that in the $n_g=1$ case [see \eq{bcrossc}],
\beq \label{bcv}
|b \cross c|^2 =   m_{\snu}^4
|\hat v \cross \hat b|^2 \,.
\eeq
Consequently, we can express
the sneutrino squared-mass splitting in the one-generation case
directly in terms of fundamental parameters of the RPV-Lagrangian in a
completely basis-independent form.

\section{The case of an arbitrary number of generations}

In this section, we obtain results for an arbitrary number of
generations.  We then explicitly exhibit the corresponding results
for $n_g=2$ and 3 generations.
The eigenvalue equation for the CP-odd scalar squared-mass matrix is
\beq \label{eigenN}
(A-\lambda) \det(C-\lambda I) + Y^{(N)}(\lambda)=0 \,,
\eeq
where $I$ is the $N\times N$ unit matrix, $N\equiv n_g$,  and
\beq \label{yndef}
Y^{(N)}(\lambda)\equiv B_i\Cof
\left[(\widetilde M^2_{\rm odd}-\lambda I)_{0i}\right]\,,
\eeq
where the sum over the repeated index $i$ is assumed implicitly.
As usual, the cofactor is defined as
$\Cof[A_{ij}]=(-1)^{i+j}\det\tilde A(i,j)$ where $\tilde A(i,j)$ is
the matrix $A$ whose $i$th row and $j$th column are removed.  
In the special case of a one-dimensional matrix, we
can define $\Cof[A_{11}]=1$.  

Let $\lambda_m^{(0)}$ ($m=1,2,\cdots,N$) be the roots of \eq{eigenN}
to zeroth order in the $B_i$, namely $\det(C-\lambda_m^{(0)} I)=0$.
For small $B_i$, we insert
$\lambda_m=\lambda_m^{(0)} + (\delta\lambda_m)_{\rm odd}$
into \eq{eigenN}.  Working to the lowest non-trivial order in
the $B_i$, we make use of \eq{detpr} to obtain 
\beq \label{oddeq2}
(\delta \lambda_m)_{\rm odd}= {Y^{(N)}(\lambda_m)
\over (A-\lambda_m) \det'(C-\lambda_m I)} \,,
\eeq 
where $\det'A$ is the product of all the non-zero 
eigenvalues of $A$.
In this analysis, we assume that there are no degenerate eigenvalues
(the degenerate case will be considered in Section~V); hence
\beq \label{detprime}
{\det}'(C-\lambda_m I)=\prod_{i\neq m} (\lambda_i-\lambda_m)\,.
\eeq
Note that we do not distinguish between $\lambda_m^{(0)}$ and
$\lambda_m$ in \eq{oddeq2}.  Since the $B_i B_j$ are the small parameters,
any distinction between the two estimates for $\lambda_m$ would yield
a result that is higher order in the product of the $B_i$.

By a similar technique, we may solve the eigenvalue equation for the 
CP-even scalar squared-mass matrix.  Noting that $\lambda_m^{(0)}$ is
the same in both the CP-odd and CP-even squared-mass computations,
we can write $\lambda_m=\lambda_m^{(0)}+(\delta\lambda_m)_{\rm even}$.
The end result is
\beq \label{eveneq2}
(\delta \lambda_m)_{\rm even}=  {Y^{(N)}(\lambda_m)(D-\lambda_m) \over 
[(D-\lambda_m)(F+A-\lambda_m)-E^2] \det'(C-\lambda_m I)}\,.
\eeq
We may evaluate the denominator of the above expression in the
RPC limit (where $B^2=0$).  In this limit,
\beqa
(D-\lambda_m)(F+A-\lambda_m)-E^2 &=&
(m_h^2-\lambda_m)(m_H^2-\lambda_m)\,,\nonumber \\
A-\lambda_m &=& m_A^2-\lambda_m\,.
\eeqa

The squared-mass difference of the $m$th sneutrino/antisneutrino pair
is denoted by $\dmsnums = (m^2_{\rm even})_m-(m^2_{\rm odd})_m$.
Plugging in the results of \eqs{oddeq2}{eveneq2}, we obtain
\beq
\dmsnums={ Y^{(N)}(\lambda_m) \over \det'(C-\lambda_m I)}
\left[{D-\lambda_m \over (m_h^2-\lambda_m)(m_H^2-\lambda_m)}-
{1 \over (m_A^2-\lambda_m)}  \right]\,.
\eeq
We may further simplify this result
by employing RPC values for any expression
that multiplies a term of order $B_i B_j$.  In particular,
we can make use of the well known tree-level MSSM Higgs results:
$\mhl^2+\mhh^2=\mha^2+\mz^2$ and $\mhh^2\mhl^2=\mha^2\mz^2\cos^2
2\beta$.  Moreover, we may take $\lambda_m=m_{\tilde \nu_m}^2$.
The end result is:
\beq \label{endresult}
\dmsnums = {m_{\tilde \nu_m}^2 m_Z^2 \sin^2 2 \beta \,
 Y^{(N)}(m_{\tilde \nu_m}^2)\over 
(m_A^2-m_{\tilde \nu_m}^2)(m_h^2-m_{\tilde \nu_m}^2)(m_H^2-m_{\tilde \nu_m}^2)
\prod_{i \ne m} (m_{\tilde \nu_i}^2-m_{\tilde \nu_m}^2)}\,.
\eeq
The small RPV parameters that govern the above expression has been
completely isolated into $Y^{(N)}$.  One additional consequence of
this result is a simple sum rule that holds for an appropriately
weighted sum of sneutrino squared-mass differences.  The sum rule and
its derivation is given in Appendix~E.

We next derive a method for computing  $Y^{(N)}(\lambda)$.  
First, we evaluate $Y^{(N)}$ for $\lambda=0$.  From \eq{yndef}, it
is straightforward to evaluate $Y^{(N)}(0)=B_i\Cof\left[(\widetilde
M_{\rm odd}^2)_{0i}\right]$.  Using \eq{modd2dnot},
\beq \label{whyzero}
Y^{(N)}(0)=-B_i B_j\Cof\left[C_{ij}\right]\,.
\eeq
Note that for $N=1$, $Y^{(1)}(\lambda)= -B^2$, independent of the
value of $\lambda$.
One can extend \eq{whyzero} for arbitrary $\lambda$.  We have found the
following recursion relation:
\beq \label{recursion}
Y^{(N)}(\lambda)= -B_i B_j\Cof\left[C_{ij}\right] 
-\lambda Y^{(N-1)}(\lambda)\,,
\eeq
with $Y^{(1)}(\lambda)=-B^2$.  The proper use of this equation
requires some care.  One must first express $Y^{(N-1)}$ covariantly in
terms of the $(N-1)$-dimensional vector $B_i$ and $(N-1)\times(N-1)$
dimensional matrix $C_{ij}$.  Then, the term $Y^{(N-1)}$ that appears
in \eq{recursion} is given by precisely the same expression [obtained
in the $(N-1)$-dimensional case], but with $B_i$ and $C_{ij}$ now
$N$-dimensional objects.  For example, 
\beq \label{why2}
Y^{(2)}(\lambda)=Y^{(2)}(0)+\lambda B^2\,,
\eeq
but in \eq{why2}, $B^2=\sum_{i=1}^N\,B_i B_i$, with $N=2$. 
 
The solution to the recursion relation [\eq{recursion}] is
\beq \label{fullwhy}
Y^{(N)}(\lambda)=\sum_{k=0}^{N-1}\,(-1)^k \lambda^k Y^{(N-k)}(0)\,.
\eeq
Again, we emphasize that the $Y^{(N-k)}$ are first obtained by an
$(N-k)$-dimensional computation.  Once these terms are expressed
covariantly in terms of $B_i$ and $C_{ij}$, the resulting expressions
for $Y^{(N-k)}$ may be used in \eq{fullwhy}, with $B_i$ and $C_{ij}$
promoted to full $N$-dimensional objects. 
Thus, for each extra generation, we need only calculate one new
invariant, $Y^{(N)}(0)$.  

We illustrate the general formulae above for the cases of $N=1$, 2 and
3 generations.  The case of $N=1$ is trivial.  Here
$Y^{(1)}(0)=-B^2$.  Using \eq{app1}, we quickly recover the result
of \eq{sqmassdiff1}.  For the case of $N=2$, we use
\eqs{whyzero}{fullwhy} to obtain:
\beq \label{why22}
Y^{(2)}(\lambda)= B^2\left[\lambda- \Tr(C)\right] +  B_iB_j C_{ij}\,.
\eeq
Using the results of Appendix~A we can express $Y^{(2)}$ directly in
terms of the model parameters:
\beq \label{why23}
Y^{(2)}(\lambda)= {1 \over v_d^2 \cos^2 \beta}
\left\{|v \cross  b|^2\,[\lambda-\Tr(\msnusnu)] +
b^2( v \cross  b)\cdot( v \cross  c)\right\} \,.
\eeq

The final result for the sneutrino squared-mass splittings in the 
two generation case is 
\beq \label{findmstwo}
\dmsnums = {4 m_{\tilde \nu_m}^2 m_Z^2 \tan^2 \beta 
\left\{ |v \cross b|^2 \,[m_{\tilde \nu_m}^2 -\Tr(\msnusnu)] +
b^2(v \cross b)\cdot(v \cross c)\right\}
\over 
v^2 (m_{\tilde \nu_n}^2-m_{\tilde \nu_m}^2) (m_A^2-m_{\tilde \nu_m}^2)
(m_h^2-m_{\tilde \nu_m}^2)(m_H^2-m_{\tilde \nu_m}^2)}\,,
\eeq
where $n \ne m$ and we have put $\lambda_m=m_{\tilde \nu_m}^2$.
We may evaluate $\Tr(\msnusnu)$ in the RPC limit:
\beq
\Tr(\msnusnu)=|b|\tan\beta+m_{\tilde \nu_1}^2+m_{\tilde \nu_2}^2\,,
\eeq
where $|b|\equiv (b_\alpha b_\alpha)^{1/2}$.
In Appendix~B, we verify that \eqs{dms}{findmstwo} agree
in the special basis. 

As in the previous section, we note that \eq{findmstwo} depends
on the derived quantity $v$.  At the expense of a somewhat more
complex result, we can re-express \eq{findmstwo} in terms of the
vectors $b$, $c$, and a new vector $c^{(2)}$ introduced in
\eq{cseries}, as shown in Appendix~C.

For $N=3$ generations, the new invariant that arises is again 
obtained from \eq{whyzero}:
\beq
Y^{(3)}(0) =
\half B^2 \left[\Tr(C^2)- [\Tr(C)]^2\right] + B_i B_j C_{ij}\,\Tr(C) -
 B_i B_j C_{ik}C_{kj} \,.
\eeq%
Following the procedure outlined above, \eqs{recursion}{why22}
yield:
\beq
Y^{(3)}(\lambda) =Y^{(3)}(0)-\lambda Y^{(2)}(\lambda)\,,
\eeq
where $Y^{(2)}(\lambda)$ is given by \eq{why23}, with $v$,
$b$ and $c$ promoted to three-dimensional vectors and $\msnusnu$
promoted to a $3\times 3$ matrix.  An explicit evaluation of 
$Y^{(3)}(0)$ is given in \eq{why3explicit}.  
Inserting the corresponding results into  \eq{endresult}, we end up
with the sneutrino squared-mass splittings in the three generation case:
\beqa \label{findmsthree}
\dmsnums &=& {-4 m_{\tilde \nu_m}^2 m_Z^2 \tan^2 \beta 
\over 
v^2 (m_{\tilde \nu_n}^2-m_{\tilde \nu_m}^2)
(m_{\tilde \nu_k}^2-m_{\tilde \nu_m}^2) (m_A^2-m_{\tilde \nu_m}^2)
(m_h^2-m_{\tilde \nu_m}^2)(m_H^2-m_{\tilde \nu_m}^2)}\nonumber \\[5pt]
&& \times\Biggl\{ |v \cross b|^2\,\Bigl[m_{\tilde \nu_m}^4 
-m_{\tilde \nu_m}^2\Tr(\msnusnu)-\half[\Tr(\msnuiv)
-[\Tr(\msnusnu)]^2\Bigr]\nonumber \\
&&\qquad + b^2(v \cross b)\cdot(v \cross c)\left[m_{\tilde \nu_m}^2
-b \cdot c -\Tr(\msnusnu)\right]+b^2 v_d^2 |b \cross c|^2\Biggr\}\,,
\eeqa
where $n\neq k \neq m$.  The traces in the RPC
limit are given by: 
\beqa
\Tr(\msnusnu)&=&|b|\tan\beta+\sum_{k=1}^3 m_{\tilde \nu_k}^2 \,,
\nonumber \\
\Tr(\msnuiv)&=&b^2\tan^2\beta+\sum_{k=1}^3 m_{\tilde \nu_k}^4 \,.
\eeqa
Again, we can check that in the special basis [see Appendix~B], 
\eqs{dms}{findmsthree} agree.
The extension of these results to four and more generations is
straightforward.

{}From the results above, we learn that if the vectors $v$ and $b$ are
parallel, then $\dmsnums=0$ for all $m$.  This is obvious for the one
and two generation cases, since we may then put $v\cross b=0$ in
\eqs{sqmassdiff1}{findmstwo}.  In the three-generation case, it is
sufficient to note that the vectors $b$ and $c$ are also parallel; it then
again follows [see \eq{findmsthree}] that $\dmsnums=0$.  In fact, if
$v$ and $b$ are parallel, then all the vectors $c^{(n)}$
[\eq{cseries}] are simultaneously parallel to $v$.  As a result,
$\dmsnums=0$ for all $m$ in the case of an arbitrary number of
generations.  Conversely, if $v$ and $b$ are not parallel, then there
must exist at least one sneutrino/anti-sneutrino pair that is split in
mass, as a consequence of the sum rule derived in \eq{sumrule}.

\section{The Degenerate Case}

So far we have assumed that the sneutrinos are non-degenerate. 
We expect this assumption to hold in any realistic model, since
the sneutrino/antisneutrino mass splittings 
are of order the neutrino masses. Thus in order for the result of the 
previous section not to hold, the flavor degeneracy has to be 
very good, namely the mass splitting between different sneutrino flavors
should be much smaller than the
neutrino mass. In any realistic model, we do not expect such 
a high degree of degeneracy. Even in models where supersymmetry breaking is
flavor blind, a mass-splitting between sneutrino flavors will
be generated via renormalization group (RG) evolution 
(from the scale of primordial
supersymmetry breaking to the electroweak scale)
that is proportional to the corresponding charged lepton masses.
Even a very small amount of running is sufficient to generate a
mass splitting that is many orders of magnitude larger than the neutrino mass.

Nevertheless, as a mathematical exercise and for completeness, we
generalize the results of the previous sections to the case of
degenerate sneutrinos.  First we give a basis-dependent
argument that explains how
one can obtain the sneutrino squared-mass splittings in the degenerate
case from the results already obtained in the non-degenerate case
without any additional calculation.  A basis-independent proof 
is relegated to Appendix~F.

Consider a case with $n_f$ sneutrinos, of which $n_d$ sneutrinos are
degenerate in mass (where $2 \le n_d \le n_f$) in the RPC
limit. Consider the $n_d$ degenerate sneutrinos and their
corresponding antisneutrinos.  Of these, $n_d-1$
sneutrino/antisneutrino pairs remain degenerate when RPV effects are
included, while one pair is spilt in mass.\footnote{This corrects a
misstatement made at the end of Section~III in \Ref{GHrpv}.}  In total,
$n_f-n_d+1$ sneutrino/antisneutrino pairs are split in mass.  The
corresponding squared-mass differences are then given by
\eq{endresult} for the $(n_f-n_d+1)$-generation case, but with all
vectors and tensors appearing in the formula promoted to $n_f$
dimensions.  The proof of this assertion is as follows. 
For the case of $n_d$ degenerate sneutrinos, the matrix $C$ [that
appears in \eqs{modd2dnot}{meven2dnot}] has $n_d$ degenerate
eigenvalues.  Thus, we are free to make arbitrary rotations within the
$n_d$ dimensional subspace corresponding to the degenerate states.  
By a suitable rotation, we can choose of basis in which only one of
the $B_i$ within the degenerate subspace is non-zero.  In
this basis the CP-odd and the CP-even squared-mass matrices 
[\eqs{modd2dnot}{meven2dnot}] separate into $(n_d-1)$ and 
$(n_f-n_d+1)$-dimensional blocks.
Clearly, the sneutrino eigenvalues in the corresponding
$(n_d-1)$-dimensional blocks are not affected by the presence of RPV terms,
while the $(n_f-n_d+1)$-dimensional block can be treated by the methods
of Section~IV.

Further generalizations, where more than one set of sneutrinos are each
separately degenerate, can also be studied.  The procedure for
computing the resulting sneutrino squared-mass differences is now clear, so we
shall not elaborate further.

\section{Discussion}

This paper provides formulae for the sneutrino/antisneutrino
squared-mass differences at tree-level in terms of basis-independent
R-parity-violating (RPV) quantities.  In contrast to the neutrino sector,
where only one tree-level neutrino mass is generated by RPV-effects, we
expect that {\it all}
sneutrino/antisneutrino squared-mass differences are 
generated at tree-level with roughly the same order of magnitude.
The sneutrino/antisneutrino mass difference is expected to be
of the same order of magnitude as the (tree-level) neutrino mass.
However, these quantities could be significantly different, as they
depend on independent RPV-parameters.
One can also analyze the case of degenerate masses for different
sneutrino flavors; although this case can only arise as a result
of a high degree of fine-tuning of low-energy parameters. 
The pattern of sneutrino/antisneutrino squared-mass differences
would provide some insight into the fundamental origin of lepton flavor
at a very high energy scale.  

The sneutrino/antisneutrino squared-mass splittings can be explored
either directly by observing sneutrino oscillation \cite{GH},
or indirectly via its effects on other lepton number violating
processes, such as neutrinoless double beta decay \cite{HKK}
and neutrino masses \cite{dpftalk,losada}.  Moreover,
the effects of tree-level sneutrino/antisneutrino squared-mass splittings
on neutrino masses are expected to be significant. 
The neutrino spectrum is determined by the relative size of the different
RPV couplings that control three sources of neutrino masses:
(i)~the tree-level mass, (ii)~the
sneutrino induced one-loop masses, and 
(iii)~the trilinear RPV induced one-loop masses \cite{GHrpv}.
Since only one neutrino acquires a tree-level mass, 
the other two mechanisms are responsible for the masses of the 
other two neutrinos. In the literature, only the trilinear RPV-induced 
one-loop masses have been considered in most studies.
In \refs{dpftalk}{losada}, it is argued that
the sneutrino-induced one-loop contributions to the neutrino masses 
are generically dominant, since the
trilinear RPV-induced one-loop masses are
additionally suppressed by a factor proportional 
to the Yukawa coupling squared.  

The results of our basis-independent formalism are useful for
comparing the two radiative neutrino mass generation mechanisms. In
particular, in models in which a theory of flavor determines the
structure of the soft-supersymmetry-breaking parameters at some high
energy scale, RG-evolution provides the connection between the
observed low-energy spectrum and the high-energy values of the
fundamental parameters of the theory \cite{enrico,RGE}.  
Basis-dependent quantities are
not renormalization-group invariant; hence the RG-evolution of
basis-independent quantities can significantly simplify the analysis.
For example, the direction of the vacuum expectation value of the
generalized slepton/Higgs scalar field is dynamically generated at
each energy scale.  Since the model parameters generically depend on
the scale, the direction of the vacuum expectation value in the
generalized lepton flavor space is scale dependent.  Clearly, in the 
basis-independent approach, such complications are avoided.
This will be the subject of a subsequent paper.

A few possible directions for future research are worth noting.
First, recall that in this paper, we assumed that CP was conserved in
the scalar sector.  If CP is violated, the required analysis is more
complicated.  Instead of diagonalizing separately CP-even and CP-odd
squared-mass matrices, one must diagonalize a single squared-mass matrix in
which the formerly CP-even and CP-odd states can mix.  Then, one must
identify the two sneutrino mass eigenstates (in the limit of small RPV
couplings).  It should be possible to extend the techniques developed
in this paper to address this more general case.  Second, in exploring
the phenomenology of sneutrino interactions (production cross-sections
and decay), one can generally assume that RPV-couplings are irrelevant
except in the decay of the lightest sneutrino state.  In that case,
new RPV-couplings enter, in particular the corresponding $\lambda$ and
$\lambda'$ parameters given in \eq{rpvsuppot}.  In the spirit of this
paper, one should also develop a basis-independent formalism to
describe the RPV sneutrino decay.  We hope to return to some of these
issues in a future work.


\acknowledgements
We thank Kiwoon Choi for helpful conversations.
YG is supported by the U.S. Department of Energy under
contract DE-AC03-76SF00515, and
HEH is supported in part by the U.S. Department of Energy
under contract DE-FG03-92ER40689.

\newpage

\appendix

\section{\boldmath{Evaluation of $Y^{(N)}(0)$ for $N=1,2$ and 3}}

Using \eq{whyzero} for $Y^{(N)}(0)$, we provide below the explicit
computation for the cases of $N=1,2$ and 3.  The computation makes use
of the definitions of $B_i$ and $C_{ij}$:
\beqa \label{beecee}
B_i &\equiv& {b_\beta X_{\beta i}\over \cos\beta}  \,,\nonumber \\[5pt]
C_{ij} &\equiv& X_{\alpha j}M^2_{\alpha\beta}X_{\beta i}\,,
\eeqa
where $M^2\equiv \msnusnu$, and the properties of the $X_{\alpha i}$
given in eqs.~(\ref{ort1})--(\ref{ort3}).  

The case of $N=1$ is very simple. 
\beq \label{app1}
Y^{(1)}(0)= -B^2 ={-1 \over \cos^2 \beta}
 b_\alpha X_{\alpha i} b_\beta X_{\beta i}
=  {-1 \over \cos^2 \beta} \left[b^2 - {(b \cdot v)^2 \over v_d^2}\right] = 
{-b^2 \over \cos^2 \beta} |\hat v \cross \hat b|^2 \,,
\eeq
where the product of cross-products, defined in \eq{xproduct}, can be
used in any number of dimensions.  

For the case of $N=2$, we compute
\beqa 
Y^{(2)}(0) &=& B_i B_j C_{ij}- B^2 \Tr(C)\nonumber \\[5pt]
&=& {1 \over \cos^2 \beta}
\left[X_{\alpha i} M^2_{\alpha \beta} X_{\beta j}
 b_\mu X_{\mu i}  b_\nu X_{\nu j}
-X_{\alpha i} M^2_{\alpha \beta} X_{\beta i}
 b_\mu X_{\mu j}  b_\nu X_{\nu j}\right] \nonumber \\[5pt]
&=&{1 \over v_d^2 \cos^2 \beta} \left\{ b^2 v_d^2 (c \cdot b) -
v_u b^2 (b \cdot v)- [b^2 v_d^2 - (b \cdot v)^2]\,\Tr (M^2) 
\right\} \nonumber \\[5pt]
&=&{1 \over v_d^2 \cos^2 \beta} 
\left\{b^2 (v \cross b) \cdot (c \cross v)- |v \cross b|^2\, \Tr (M^2)\right\}
\,. 
\eeqa
Note that the resulting expression has simplified considerably after
introducing the vector $c$ [defined in \eq{cdef}].

The case of $N=3$ is more involved.
\beq
Y^{(3)}(0) =\half B^2 \left[\Tr(C^2)- [\Tr(C)]^2\right] 
+ B_i B_j C_{ij}\,\Tr(C) - B_i B_j C_{ik}C_{kj} \,.
\eeq
We calculate separately the two terms above.  First, $B^2$ is obtained
from \eq{app1} and
\beqa \label{app2}
\Tr(C^2)-[\Tr(C)]^2&=&
\left[X_{\alpha i} M^2_{\alpha \beta} X_{\beta j}
X_{\mu j} M^2_{\mu \nu} X_{\nu i}
-X_{\alpha i} M^2_{\alpha \beta} X_{\beta i}
X_{\mu j} M^2_{\mu \nu} X_{\nu j}\right] \nonumber \\
&=&
\Tr(M^4)-[\Tr(M^2)]^2 - {2 v_u \over v_d^2} 
\left[  v_u b^2-(b \cdot v)\,\Tr(M^2)\right]\,. 
\eeqa
%
Next, we evaluate
\beqa \label{app3}
&&[B_i B_j C_{ij}\,\Tr(C) - B_i B_j C^2_{ij}]\cos^2\beta \nonumber \\ [5pt]
&&\qquad =
X_{\alpha i} M^2_{\alpha \beta} X_{\beta i}
X_{\mu j} M^2_{\mu \nu} X_{\nu k}  b_\sigma X_{\sigma k}  b_\rho X_{\rho j}-
X_{\alpha i} M^2_{\alpha \beta} X_{\beta j}
X_{\mu j} M^2_{\mu \nu} X_{\nu k}  b_\sigma X_{\sigma k}  b_\rho X_{\rho i}
 \nonumber \\[5pt]
&&\qquad =
 M^2_{\mu \nu} b_\mu b_\nu\, \Tr(M^2) - 
M^2_{\alpha \mu} M^2_{\mu \nu} b_\nu b_\alpha
-{v_u \over v_d^2} \left[
 b^2 (b \cdot v)\, \Tr(M^2) - (b \cdot v) M^2_{\mu \nu} b_\nu b_\alpha
\right]\nonumber \\[5pt]
&& \qquad\qquad\qquad
+{v_u \over v_d^2} \left[
v_u b^4 - b^2 (b \cdot v)\,\Tr(M^2)\right]
+ {v_u \over v_d^4} \left[ (b \cdot v)^3 \,\Tr(M^2)
-  v_u b^2 (b \cdot v)^2
\right] \,. 
\eeqa
The above result can be simplified further.  First, the last
two terms can be combined by noting that
\beq
v_u b^4- b^2 (b \cdot v)\,\Tr(M^2)
+ {1 \over v_d^2} \left[ (b \cdot v)^3\,\Tr(M^2) 
-  v_u b^2 (b \cdot v)^2\right] = {|v \cross b|^2\over v_d^2}    
\left[v_u b^2-(b \cdot v)\,\Tr(M^2)\right] . 
\eeq
This term will end up canceling a similar term in \eq{app2}.

At this point, it is convenient to re-express some of the terms of
\eq{app3} in terms of the vector $c$.  First, we observe that
\beqa
v_d^2 M^2_{\mu \nu} b_\mu b_\nu\,\Tr(M^2) - v_u b^2 (b \cdot v)\, \Tr(M^2)
&=&  b^2 \left[(b \cdot c) v_d^2 - (b \cdot v) (c \cdot v)\right]\, \Tr(M^2)
\nonumber \\[5pt]
&=&  b^2 (v \cross c)\cdot(v \cross b)\, \Tr(M^2) \,. 
\eeqa
In deriving the above result, we noted that 
$M^2_{\mu \nu} b_\nu v_\nu = v_u b^2 = b^2 (v \cdot c)$ 
[using \eqs{veveq}{cdef} and the fact that $\msnusnu$ is a symmetric matrix], 
which implies that 
\beq \label{ctanb}
v_u=v\cdot c\,.
\eeq
Second,
\beqa
v_d^2 M^2_{\alpha \mu} M^2_{\mu \nu} b_\nu b_\alpha 
-v_u (b \cdot v) M^2_{\mu \nu} b_\nu b_\nu 
&=& b^4 c^2 v_d^2 - v_u b^2 (b \cdot v)(b \cdot c) \nonumber \\[5pt]
&=& b^2\left[
v_d^2|b \cross c|^2 + (b \cdot c) (v \cross b) \cdot (v \cross c)\right]\,.
\eeqa
Collecting all of the above results, the final expression is quite compact:
\beqa \label{why3explicit}
Y^{(3)}(0) &=& 
{1 \over v_d^2\cos^2 \beta} 
\left\{\half
|v \cross b|^2 \left[\Tr(M^4)-[\Tr(M^2)]^2 \right]
+ b^2 v_d^2 |b \cross c|^2 \right. \nonumber \\[5pt]
&&\qquad + \left.
b^2 (v \cross c)\cdot(v \cross b)\left[\Tr(M^2) -(b \cdot c)\right] \right\}\,.
\eeqa

\section{Sneutrino Squared-Mass Splitting Formulae in the 
Special Basis}

We define the special basis in which
$v_m=0$ ({\it i.e}, the neutral
scalar vacuum expectation values determines the definition of the
down-type Higgs field) and the matrix $\msnusnuij$ 
[which is the $3\times 3$ block sub-matrix of
$\msnusnuab$ defined in \eq{treelevelsnumasses}] is diagonal.
As in Appendix~A we define $M^2\equiv \msnusnu$.

In the special basis, 
one can use eqs.~(\ref{veveq}), (\ref{cdef}) and
(\ref{ctanb}) to obtain the following relations:
\beq \label{cmtwo}
c_0=\tan\beta\,, \qquad 
M^2_{00}=b_0 c_0\,, \qquad
M_{0i}^2=b_i \tan \beta\,, \qquad
c_i = {b_i (M^2_{00}+M^2_{ii}) \over b^2}\,.
\eeq  
Using these results, it follows that
\beqa \label{useres}
|v\cross b|^2 &=& v_d^2\sum_i b_i^2\,, \\[5pt]
b^2(v\cross b)\cdot (v\cross c) &=& v_d^2\left[M_{00}^2\sum_i b_i^2
+\sum_i M_{ii}^2b_i^2\right]\,.\label{useres2}
\eeqa
We will also need a similar expression for $|b\cross c|^2$.  First, we
note that the last relation of \eq{cmtwo} implies
\beq
c^2 = c_0^2 + {1\over b^4}\sum_i b_i^2 (M^2_{00}+M^2_{ii})^2  \,, \qquad
b \cdot c = M^2_{00} + {1\over b^2}
\sum_i b_i^2 (M^2_{00}+M^2_{ii}) \,.
\eeq
These results can be used to obtain:
\beq \label{t3term1}
|b \cross c|^2 = 
b^2c^2 - (b \cdot c)^2 = 
{1\over b^2}\sum_i (b_i M^2_{ii})^2  + {\cal O}(b_i^4) \,.
\eeq

We now turn to the specific cases.  For the case of one 
generation, the basis-independent result is given in \eq{sqmassdiff1}.
In the special basis, \eq{useres} yields
\beq
b^2|\hat v \cross \hat b|^2 = b_1^2.
\eeq 
Inserting this result into \eq{sqmassdiff1}, we immediately obtain \eq{dms}.

In the two generation case,
the basis-independent result is given in \eq{findmstwo}.
In the special basis, \eqs{useres}{useres2} yield
\beqa \label{twogen1}
b^2(v \cross b)\cdot (v \cross c) &=&
v_d^2\left[M_{00}^2(b_1^2 + b_2^2) + M_{11}^2 b_1^2 + M_{22}^2 b_2^2
\right]\,, \\[5pt]
[M^2_{ii} -\Tr(M^2)] |v \cross b|^2 &=&
-v_d^2\left(M^2_{00}+M^2_{jj}\right) (b_1^2 + b_2^2)\,,\label{twogen2}
\eeqa%
for the two cases of $i=1$, $j=2$ and $i=2$, $j=1$, respectively.
Adding the above two equations, one finds
\beq
[M^2_{ii} -\Tr(M^2)] |v \cross b|^2 +
b^2(v \cross b)\cdot (v \cross c) =
(M_{ii}^2 -M_{jj}^2)b_i^2 v_d^2 \,.
\eeq
Working to leading order in the RPV-parameters $b_i^2$, we may set
the diagonal elements of $M^2$ to their RPC values, 
$M_{ii}^2=m^2_{\tilde\nu_i}$.
Plugging the result into \eq{findmstwo}, one again
recovers \eq{dms}.

In the three generation case,
the basis-independent result is given in \eq{findmsthree}.  Again,
it is sufficient to work to leading order in the $b_i^2$.
Then, one finds that in the special basis,
\beqa
[M_{11}^2]^2-M_{11}^2 \Tr(M^2) -\half[\Tr(M^4)-[\Tr(M^2)]^2]
&=& M_{00}^2\left[M^2_{22}+M^2_{33}\right]+
M^2_{22} M^2_{33}+{\cal O}(b_i^2)\,,\nonumber \\[5pt]
M_{11}^2-b \cdot c -\Tr(M^2) &=& -M_{22}^2-M_{33}^2+{\cal O}(b_i^2)\,.
\eeqa
Using these results and those of eqs.~(\ref{useres}), (\ref{useres2})
and (\ref{t3term1}), we end up with
\beqa
&&\left\{[M_{11}^2]^2-M_{11}^2\Tr(M^2)-\half[\Tr(M^4)
-[\Tr(M^2)]^2\right\} |v \cross b|^2 +b^2 v_d^2 |b \cross c|^2
\nonumber \\[5pt]
&&\qquad + b^2(v \cross b)\cdot(v \cross c)\left[M_{11}^2
-b \cdot c -\Tr(M^2)\right]=
v_d^2\, b_1^2\, (M_{11}^2-M_{22}^2) (M_{11}^2-M_{33}^2)\,.
\nonumber
\eeqa
Two additional equations can be generated by permuting the
indices 1,2 and 3.  Finally, setting $M_{ii}^2=m^2_{\tilde\nu_i}$
and plugging the result into \eq{findmsthree}, one 
confirms \eq{dms} for the third time.

\section{\boldmath{How to eliminate $\lowercase{v}$ in favor of other vectors}}

Our final expressions for the sneutrino squared mass differences
depend on basis-independent products of vectors, $v$, $b$, $c,\ldots\,$,
and traces of powers of $\msnusnu$.  However, the vector
$v$ is not a fundamental parameter of the model, but a derived
parameter which arises as a solution to \eq{veveq}.  With some
manipulation, it is possible to eliminate $v$ in favor of the other
vectors (which correspond more directly to the fundamental
supersymmetric model parameters, namely
$b$ and a series of vectors obtained by
multiplying $b$ some number of times by $\msnusnu$).
In this appendix, we illustrate the procedure in the case of the one
and two generation models.  

In the one generation model, $\msnusnu$ is a $2\times 2$ matrix.
Consider an arbitrary
$2 \times 2$ matrix $A$ and its characteristic equation
$\det(A-\lambda I)=0$.  Since any matrix satisfies its own
characteristic equation, we obtain\footnote{We henceforth suppress the
obvious factors of the identity matrix $I$.}
\beq
A^2 - A \Tr(A) + \det (A) = 0\,,
\eeq
which after multiplication by $A^{-1}$ yields
\beq \label{mat2}
A^{-1}= {\Tr(A)-A \over \det (A)}\,.
\eeq
Using \eq{mat2} we can express $|v \cross b|^2$ in terms of  $|b \cross c|^2$.
Let $A\equiv\msnusnu$, and use \eqs{veveq}{cdef} to obtain
\beq \label{vee}
v= v_u A^{-1} b = {v_u\left [b \Tr(A)-b^2 c\right] \over \det (A)}\,,
\eeq
We now substitute \eq{vee} for $v$ in $|v \cross b|^2\equiv b^2 v_d^2 
- (v \cdot b)^2$.
\beqa
b^2 v_d^2 &=&{b^4 v_u^2 \over [\det (A)]^2}\left\{
[\Tr(A)]^2 - 2(b \cdot c)\Tr(A) + b^2 c^2\right\}\,, \nonumber \\[5pt]
(v \cdot b)^2 &=& {b^4 v_u^2 \over [\det (A)]^2}\left\{
[\Tr(A)]^2 - 2(b \cdot c)\Tr(A) + (b \cdot c)^2\right\}\,.
\eeqa
Subtracting these two equations, we end up with
\beq \label{vcrossb}
|v \cross b|^2 = {b^4 v_u^2 \over [\det (A)]^2} |b \cross c|^2\,.
\eeq
Since $|b \cross c|^2$ is the small RPV parameter, we may evaluate 
$\det(A)$ in the RPC limit.  Using \eq{veveq} in the RPC limit,
$[\det (A)]^2 = m_{\snu}^4 \, b^2 \tan^2 \beta$.
The end result is
\beq \label{bcrossc}
|b \cross c|^2 = m_{\snu}^4|\hat v \cross \hat b|^2\,.
\eeq

In the two generation model, $\msnusnu$ is a $3\times 3$ matrix.
The procedure again employs the characteristic equation.  For an
arbitrary $3\times 3$ matrix,
\beq
A^{-1}= {A^2-A\Tr(A)+\Sym A \over \det (A)}\,,
\eeq
where\footnote{To prove \eq{sym2}, simply note that $
\Tr(A^2)=\sum_k \lambda_k^2$.} 
\beq \label{sym2}
\Sym (A) \equiv \sum_{i<j} \lambda_i \lambda_j\ 
=\half\left\{\left[\Tr(A)\right]^2-\Tr(A^2)\right\}\,,
\eeq
where $\lambda_j$ are the eigenvalues of $A$.  We can again solve for
$v$ following the method used in the one-generation case:
\beq
v = v_u A^{-1} b = {v_u\left[b^2 d -b^2 c\, \Tr(A)+b \,\Sym (A)
\right] \over \det (A)}\,,
\eeq
where we have defined $d \equiv c^{(2)}=Ac$.  After some algebra, we 
obtain\footnote{Observe that the result for $|v\cross b|$ depends on
the number of generations, {\it i.e.} the dimension of the matrix $A$
[compare \eqs{vcrossb}{vcrossb2}].}
\beq \label{vcrossb2}
|v \cross b|^2 = {b^4 v_u^2 \over [\det (A)]^2} \left[ 
|b \cross d|^2 + |b \cross c|^2[\Tr(A)]^2 
-2(b \cross c)(b \cross d)\Tr(A)\right]\,,
\eeq
and
\beqa
(v \cross b) \cdot (v \cross c) &=& 
 {b^2 v_u^2 \over [\det (A)]^2} \left[
(d \cross b)\cdot(d \cross c)[b^2-\Sym (A)]\right.\nonumber \\[6pt] 
&&\qquad \left. + b^2(c \cross d)\cdot(c \cross b) \Tr(A)
+ |b \cross c|^2\, \Tr(A)\, \Sym (A)\right]\,. 
\eeqa
It is easy to evaluate the three invariants in the RPC 
limit:\footnote{Since $v\cdot c=v_u$, it follows that in the RPC limit, 
$|c|=\tan\beta$.} 
\beqa
\Tr(A)&=&|b|\tan\beta+m_1^2+m_2^2\,,\nonumber \\ 
\det(A)&=&m_1 m_2 |b|\tan\beta\,,\nonumber \\
\Sym (A)&=& m_1^2 m_2^2 +|b|\tan\beta(m_1^2+m_2)^2\,,
\eeqa
where $m_i^2\equiv m^2_{\tilde\nu_i}$.

Inserting the above results into \eq{findmstwo} yields the desired
result.  Further algebraic manipulations of the resulting expression
do not lead to a particularly simple result. 
 
\section{\boldmath{Evaluation of 
$\det[A-(\lambda_{\lowercase{m}}-\epsilon)I]$}}

Consider a general $N\times N$ matrix $A$, with eigenvalues $\lambda_k$.
Then,
\beq
\det(A-\lambda I)=\prod_k^N (\lambda_k-\lambda)\,,
\eeq
where the product is taken over all $N$ eigenvalues (some of which might
be degenerate).  First, suppose that there are no degenerate
eigenvalues.  If $\lambda_m$ is one of the eigenvalues and
$\epsilon\ll 1$, then
\beqa \label{detpr}
\det[A-(\lambda_m-\epsilon)I] &=&\epsilon\prod_{i\neq m}(\lambda_i-\lambda_m)
+{\cal O}(\epsilon^2)\nonumber \\[5pt]
&=&\epsilon\,{\det}'(A-\lambda_m I)+{\cal O}
(\epsilon^2) \,,
\eeqa
where $\det'M$ is the product of all the non-zero 
eigenvalues of $M$ [see \eq{detprime}].

The case of degenerate eigenvalues is easily handled.
We can still use \eq{detprime} if it is understood that all terms in
which $\lambda_i$ is equal to the degenerate eigenvalue 
are omitted from the product.  If
$\lambda_d$ is an eigenvalue which is $n_d$-fold degenerate,   
then \eq{detpr} is generalized to
\beq \label{detpr-deg}
\det[A-(\lambda_d-\epsilon)I]=\epsilon^{n_d}\,{\det}'(A-\lambda_d I)+{\cal O}
(\epsilon^{n_d+1}) \,.
\eeq

In Section IV and Appendix~F, we have employed these results with
$\epsilon=-\delta\lambda_m$ and $\epsilon=-\delta\lambda_d$, respectively.

\section{Sneutrino Squared-Mass Splitting Sum Rules}

In the case of $N$ sneutrino generations, one can calculate the
corresponding sneutrino squared-mass splittings.   In the case of
non-degenerate tree-level sneutrino masses, the squared-mass
splittings were obtained in \eq{endresult}.  In the case of degenerate
masses, one employs the modified results according to the discussion given
in Section~V and Appendix~F.  We then find the following interesting
sum rule:
\beq \label{sumrule}
\sum_{m=1}^{N} 
{v^2 (m_A^2-m_{\tilde \nu_m}^2)(m_h^2-m_{\tilde \nu_m}^2)
(m_H^2-m_{\tilde \nu_m}^2)
\over 4 m_Z^2\tan^2\beta}{\dmsnums\over m_{\tilde \nu_m}^2} 
= -|v \times b|^2\,.
\eeq

We shall prove this result for the non-degenerate case.  Using
\eqs{endresult}{app1}, we see that \eq{sumrule} is equivalent to the
following result:
\beq \label{ysum}
\sum_{m=1}^{N} {Y^{(N)}(\lambda_m)\over
\prod_{i\ne m}(\lambda_i-\lambda_m)}=Y^{(1)}(0)\,.
\eeq
To prove \eq{ysum}, we insert the expansion for $Y^{(N)}(\lambda_m)$
[\eq{fullwhy}] into \eq{ysum}, and make use the following identity:
\beq \label{identity}
S_{N,k}\equiv \sum_{m=1}^{N} 
{\lambda_m^k\over\prod_{i \ne m} (\lambda_m-\lambda_i)}
=\cases{0\,,& $k=0,1,\ldots,N-2$\,,\cr 1\,, & $k=N-1$\,,\cr}
\eeq
where all the $\lambda_m$ are assumed to be 
distinct.\footnote{Note that in
\eq{identity}, the sign of the factors $\lambda_m-\lambda_i$ is
reversed compared to \eq{ysum}.  Thus, an extra factor of $(-1)^{N-1}$
is generated which cancels with the corresponding sign in front of 
$Y^{(1)}$ in \eq{fullwhy}.}

\Eq{identity} is established as follows.
Let $f(x)=(x-\lambda_1)(x-\lambda_2)\cdots(x-\lambda_N)$,
where the $\lambda_m$ are distinct.  Consider the
resolution of $x^{k+1}/f(x)$ into partial fractions
(where $k$ is an integer such that $0\leq k\leq N-2$):
\beq \label{partfrac}
{x^{k+1}\over f(x)}=\sum_{m=1}^N {A_m\over x-\lambda_m}\,.
\eeq
Combining denominators, it follows that:
\beq \label{partfracnum}
x^{k+1}=\sum_{m=1}^N A_m\prod_{i\ne m}(x-\lambda_i)\,.
\eeq
The right hand side of \eq{partfracnum} is a polynomial of degree
$N-1$ or less.  Since this must be an identity for all $x$, 
we can solve for each coefficient $A_m$ separately 
by setting $x=\lambda_m$:
\beq 
A_m={\lambda_m^{k+1}\over \prod_{i\ne m}(\lambda_m-\lambda_i)}\,.
\eeq
Inserting this result into \eq{partfrac} and setting $x=0$ yields
\eq{identity} for the case of $0\leq k\leq N-2$.  The case of $k=0$
where one of the $\lambda_m$ vanishes must be treated separately,
although it is easy to show that the end result is unchanged.  Thus,
$S_{N,k}=0$ for $0\leq k\leq N-2$.

To derive \eq{identity} in the case of $k=N-1$, we
set $k=N-2$ in \eq{partfracnum}.  On the right hand side of
\eq{partfracnum}, we note that the term proportional to $x^{N-1}$
arises simply by setting the $\lambda_i=0$.  It follows that
$\sum_{m=1}^N\,A_m=1$ (for $k=N-2$) which is precisely equivalent to
$S_{N,N-1}=1$ and the proof is complete. 

Finally, we note a useful recursion relation satisfied by the $S_{N,k}$.
Multiply the $m$th term of \eq{identity} by $(\lambda_m-\lambda_{N+1})/
(\lambda_m-\lambda_{N+1})$.  One immediately deduces that
relation:
\beq \label{Srecursion}
S_{N,k}=S_{N+1,k+1}-\lambda_{N+1} S_{N+1,k}\,.
\eeq
The boundary conditions for the recursion relation are:
$S_{N,0}=0$ for $N\geq 2$ (which is a consequence of the proof given
above), and $S_{1,0}=1$.\footnote{The condition $S_{1,0}=1$ formally
defines the sum [\eq{identity}] in the case of $N=1$.  Alternatively,
one can check by explicit evaluation that $S_{2,1}=1$.  Thus, we see
that the assigned definition of $S_{1,0}$ is consistent.
Note that one can similarly define $S_{1,k}=\lambda^k_1$.}
It follows that $S_{N,k}=0$ for $1\leq k\leq N-2$.
Choosing $k=N-1$ in \eq{Srecursion} then yields $S_{N+1,N}=S_{N,N-1}$;
it follows that $S_{N+1,N}=1$ for all $N\geq 1$.  Finally, it is easy
to increase $k$ further.  For example, \eq{Srecursion} 
implies that $S_{N+1,N+1}=S_{N,N}+\lambda_{N+1}$.  It follows that
$S_{N,N}=\sum_{i=1}^N \lambda_i$, and so on.
 
The degenerate case can be treated as a limiting case of the
non-degenerate results obtained above.
In the final analysis, we find that \eq{sumrule}
applies in general, and serves as a useful check of our results.

\section{Basis-Independent Treatment of the Degenerate Case}

The degenerate case for $n_f$ flavors was treated in Section~V.  If $n_d$
sneutrino/antisneutrino pairs are degenerate in mass in the RPC limit,
then when RPV-effects are incorporated, one finds that $n_d-1$ pairs
remain degenerate, while $n_f-n_d+1$ sneutrino/antisneutrino pairs are
split in mass.  The squared-mass splittings of the latter can be
obtained from the corresponding formulae of the 
non-degenerate $n_f-n_d+1$ flavor case.

In this appendix, we briefly sketch the required steps of a proof
that generalizes the basis-independent results of Section~IV. 
Consider first the
squared-mass matrix of the CP-odd scalars [\eq{modd2dnot}].  The
characteristic equation, \eq{eigenN}, is still valid in the case of
degenerate sneutrinos.  First we consider the quantity
$Y^{(N)}(\lambda)$ [\eq{yndef}].  Suppose that the matrix $C$ which
appears in $\widetilde M^2_{\rm odd}$ has an eigenvalue $\lambda_d$
that is $n_d$-fold degenerate (with the remaining eigenvalues of $C$
distinct).  We assert that the following formula holds:
\beq \label{yndeg}
Y^{(N)}(\lambda)= (\lambda_d-\lambda)^{n_d-1}
Y_{\rm deg}^{(N-n_d+1)}(\lambda)\,,
\eeq
where $Y_{\rm deg}^{(N-n_d+1)}(\lambda)$ is obtained as follows.
First, one evaluates $Y^{(N-n_d+1)}(\lambda)$ as in Section~IV, and
expresses the result covariantly in terms of the vector $B_i$ and the 
matrix $C_{ij}$.  Next, these quantities are reinterpreted as
$N$-dimensional objects.   Finally, all traces that appear in the
result are replaced by:
\beq \label{trprime}
\Tr^{\,\prime}\, C^n\equiv\Tr\, C^n-(n_d-1)\lambda_d^n\,.
\eeq

Consider first the effect of the RPV terms on the non-degenerate 
sneutrinos.  Then the analysis of Section~IV can be used, and we
obtain \eq{endresult} for the squared-mass splitting of
sneutrino/antisneutrino pairs.  If we now insert \eq{yndeg} for
$Y^{(N)}(m^2_{\tilde\nu_m})$, we see that we obtain a new formula
which has the same form as \eq{endresult}, with the following
modifications: (i) $Y^{(N)}(m^2_{\tilde\nu_m})$ is replaced by
$Y_{\rm deg}^{(N-n_d+1)}(m^2_{\tilde\nu_m})$; and (ii) the product
that appears in the denominator of \eq{endresult} is modified to
$\prod'_{i\ne m} (m_{\tilde \nu_i}^2-m_{\tilde \nu_m}^2)$, 
where the prime indicates that degenerate squared-masses appear only
once in the product.  To obtain a covariant expression for 
$Y_{\rm deg}^{(N-n_d+1)}(m^2_{\tilde\nu_m})$, we first obtain the 
expression of $Y^{(N-n_d+1)}(m^2_{\tilde\nu_m})$
in terms of the various vectors ($v$, $b$, $c,\ldots$) and 
traces of powers of $M^2$
using the results of Section~IV and Appendix~A.
The resulting expression can then be used for 
$Y_{\rm deg}^{(N-n_d+1)}(m^2_{\tilde\nu_m})$ by replacing
$\Tr\, M^{2n}$ with $\Tr^{\,\prime}\,M^{2n}$ [the latter is defined
by replacing $C$ with $M^2$ in \eq{trprime}] and interpreting
all the vectors and matrices as $N$-dimensional objects.

Finally, consider the effect of the RPV terms on the $n_d$ degenerate
sneutrinos.  Now, we must return to \eq{eigenN} and insert 
$\lambda=\lambda_d^{(0)} + (\delta\lambda_d)_{\rm odd}$.  
Working to the lowest non-trivial order in
the $B_i$, we make use of \eqs{yndeg}{detpr-deg} to obtain 
\beq \label{oddeq2deg}
\left[(\delta \lambda_d)_{\rm odd}\right]^{n_d}= 
{Y_{\rm deg}^{(N-n_d+1)}(\lambda_d)
\left[(\delta \lambda_d)_{\rm odd}\right]^{n_d-1}
\over (A-\lambda_d) \det'(C-\lambda_d I)}\,.
\eeq 
The solution to this equation has $n_d-1$ degenerate solution, 
$(\delta \lambda_d)_{\rm odd}=0$, and one non-degenerate solution
for $(\delta \lambda_d)_{\rm odd}$ which has the same form as
\eq{oddeq2} for the $(N-n_d+1)$-dimensional problem.  As described
above, we can make use of the relevant covariant expressions 
obtained in Section~IV and Appendix~A by
replacing $\Tr$ with $\Tr^{\,\prime}$ and interpreting
all the vectors and matrices as $N$-dimensional objects.
The end result is that $n_d-1$
sneutrino/antisneutrino pairs remain degenerate, while one of the
original degenerate pairs is split according to the
$N-n_d+1$-dimensional version of \eq{endresult} [with all vectors and
tensors promoted to $N$-dimensional objects].  One can also check that the
sum rule obtained in Appendix~E for sneutrino squared-mass differences
(appropriately weighted) applies even when there are degenerate
sneutrino masses.

{\tighten

}

\end{document}